\documentclass[12pt]{article}
\usepackage{amssymb}
\usepackage{epsfig}
\usepackage{float}
\usepackage{graphicx}
\usepackage{amsmath}
\newcommand{\nua}[1]{\ensuremath{\rlap
           {\kern-2.5pt\ensuremath
           {\overset{\scriptscriptstyle(-)}{\phantom{\nu}}}}
           {\ensuremath{{\nu}_{#1}}}}}

\begin{document}
\begin{center}
{\bf  On the  Origin of Majorana Neutrino Masses}\footnote{ A talk at the 5-th International Conference on Particle Physics and Astrophysics, Moscow, Russia, 5-9 October 2020}
\end{center}
\begin{center}
{ \bf S. M. Bilenky}
\end{center}
\begin{center}
{\em  Joint Institute for Nuclear
Research, Dubna, R-141980, Russia\\}{\em TRIUMF
4004 Wesbrook Mall,
Vancouver BC, V6T 2A3
Canada\\}
\end{center}
\begin{abstract}
In the introductory part we briefly consider  basics of neutrino oscillations and convenient phenomenology of neutrino oscillations in vacuum. Main part of this report is dedicated to a
discussion of a plausible BSM scenarios of neutrino mass generation, based on assumptions of massless left-handed SM neutrinos and violation of the total lepton number. It is stressed that search for light sterile neutrinos and neutrinoless double $\beta$-decay could provide a crucial tests of this scenarios.
\end{abstract}

\section{Introduction}
One of the most important recent discoveries in  particle physics was discovery of neutrino oscillations in the atmospheric Super-Kamiokande \cite{Fukuda:1998mi}, solar SNO \cite{Ahmad:2002jz} and  the reactor KamLAND experiment \cite{Eguchi:2002dm} (1998-2002). In 2015 the Nobel Prize was awarded to T. Kajita and A. McDonald ``for the discovery of neutrino oscillations, which shows that neutrinos have masses". Small neutrino masses, driving neutrino oscillations, is the only evidence (in particle physics) of an existence of a new beyond the Standard Model physics.

Origin of neutrino masses, which are many orders of magnitude smaller than quark and lepton masses, is the major open problem of neutrino physics.

In the first part of this talk I will consider phenomenon of neutrino oscillations. In the second part I will discuss a possible (and plausible) origin of neutrino masses and mixing.

\section{Basics of Neutrino Oscillations}
Idea of neutrino oscillations was put forward by B.Pontecorvo in Dubna in 1957-58 \cite{Pontecorvo:1957qd}. This idea was further  developed by B.Pontecorvo and V.Gribov (1969)\cite{Gribov:1968kq} and B.Pontecorvo and S.Bilenky (1975-1989)(see, for example, \cite{Bilenky:1978nj}). Idea of flavor neutrino mixing was proposed in \cite{Maki:1962mu}.

We know from experiments on the investigation of the "invisible" decay $Z^{0}\to \nu_{l}+\bar\nu_{l}$ (LEP, SLC) that three flavor left-handed neutrinos $\nu_{e},\nu_{\mu},\nu_{\tau}$ (and right-handed antineutrinos) exist in nature. The left-handed fields of flavor neutrinos $\nu_{lL}$~~$(l=e,\mu,\tau)$ enter into the Standard Model CC and NC weak interaction
\begin{equation}\label{CC}
    \mathcal{L^{CC}_{I}}=-\frac{g}{2\sqrt{2}}j^{CC}_{\alpha}W^{\alpha}+
\mathrm{h.c.},\quad j^{CC}_{\alpha}=
2\sum_{l=e,\mu,\tau}\bar\nu_{lL}\gamma_{\alpha}l_{L}
\end{equation}
and
\begin{equation}\label{NC}
    \mathcal{L^{NC}_{I}}=-\frac{g}{2\cos\theta_{W}}j^{NC}_{\alpha}
    Z^{\alpha}, \quad j^{NC}_{\alpha}=
\sum_{l=e,\mu,\tau}\bar\nu_{L}\gamma_{\alpha}\nu_{L}.
\end{equation}
From the observation of neutrino oscillations follows that {\em flavor neutrino fields are mixed}
\begin{equation}\label{mix}
\nu_{lL}(x)=\sum^{3}_{i=1}U_{li}~\nu_{iL}(x).
\end{equation}
Here $\nu_{i}(x)$  is the field of neutrino (Dirac or Majorana) with mass $m_{i}$ and $U$ is $3\times3$ PMNS mixing matrix.

It follows from (\ref{mix}) that if all neutrino mass-squared differences are small, states of the flavor neutrinos with definite momentum, produced in the decays $\pi^{+}\to \mu^{+}+\nu_{\mu}$ (accelerator and atmospheric neutrinos),
 $(A,Z)\to (A,Z+1)+e^{-}+\bar\nu_{e}$ (reactor antineutrinos) etc are given by
\begin{equation}\label{states}
  |\nu_{l}\rangle=\sum^{3}_{i=1}U^{*}_{li}~|\nu_{iL}\rangle~~
  (l=e,\mu,\tau),
\end{equation}
where $|\nu_{i}\rangle$ is the state of neutrino with mass $m_{i}$, momentum $\vec{p}$ and energy $E_{i}\simeq p +\frac{m^{2}_{i}}{2E}$~ ($p^{2}\gg m^{2}_{i}$).

If at $t=0$ flavor neutrino $\nu_{l}$ was produced, the state of neutrino at the time  $t$  will be  a coherent superposition
\begin{equation}\label{transit}
|\nu_{l}\rangle_{t}=\sum_{i}U^{*}_{li}~e^{-iE_{i}t}|\nu_{iL}\rangle    =\sum_{l'}\mathcal{A}(\nu_{l}\to \nu_{l'})|\nu_{l'}\rangle.
\end{equation}
Here
\begin{equation}\label{transit1}
\mathcal{A}(\nu_{l}\to \nu_{l'})=e^{-ipL}\sum_{i}U_{l'i}
e^{-i\frac{m^{2}_{i}L}{2E}}U^{*}_{li}
\end{equation}
is the amplitude of transition $\nu_{l}\to \nu_{l'}$ during time $t$, $L\simeq t$ is a source-detector distance.

The amplitude $\mathcal{A}(\nu_{l}\to \nu_{l'})$ is {\em a coherent sum} of products of  amplitudes (of transition $(\nu_{l}\to\nu_{i})$ ($U^{*}_{li}$)) (of propagation in the state $\nu_{i}$ ($ e^{-iE_{i}t}$) ) (of transition $(\nu_{i}\to\nu_{l'})$~($U_{l'i}$)).

Neutrino oscillations in vacuum are the result of  {\em interference
between different $i$-amplitudes}. Taking into account the unitarity of the mixing matrix $U$ and the arbitraness of a common phase from (\ref{transit}) and (\ref{transit1}) we find a convenient expression for the transition probability (see \cite{Bilenky:2015xwa})
\begin{equation}\label{Probabil2}
\mathrm{P}(\nu_{l}\to \nu_{l'})= |\sum^{3}_{i=1}U_{l'i}
e^{-i\frac{m^{2}_{i}L}{2E}}U^{*}_{li}|^{2}=|\delta_{l'l}
-2i\sum_{i\neq r}e^{-i\Delta_{ri}}U_{l'i}U^{*}_{li}\sin\Delta_{ri}|^{2}.
\end{equation}
Here
\begin{equation}\label{Probabil1}
 \Delta_{ri}=\frac{\Delta m^{2}_{ri}L}{4E},\quad\Delta m^{2}_{ik}= m^{2}_{k}- m^{2}_{i}
\end{equation}
and $r$ is an arbitrary, fixed index.

The second term of (\ref{Probabil2}) describes neutrino oscillations. We see from this expression that neutrino oscillations take place if
\begin{itemize}
  \item at least one neutrino neutrino mass-squared difference is different from zero;
  \item there is a neutrino mixing ($U\neq I$).
 \end{itemize}
Let us consider the simplest case of two flavors, say $\mu$ and $\tau$. Choosing $r=1$  from (\ref{Probabil2}) we have
\begin{equation}\label{Probabil3}
\mathrm{P}(\nu_{l}\to \nu_{l'})=|\delta_{l'l}
-2ie^{-i\Delta_{12}}U_{l'2}U^{*}_{l2}\sin\Delta_{12}|^{2}.
\end{equation}
For two flavors the mixing matrix has the following general form
\begin{equation}\label{Probabil4}
U=\left(
\begin{array}{c c}
\cos\theta & \sin\theta \\
-\sin\theta   & \cos\theta \\
 \end{array}
\right)
\end{equation}
From (\ref{Probabil3}) and (\ref{Probabil4}) we find the standard two-neutrino appearance and disappearance transition probabilities
\begin{equation}\label{Probabil5}
\mathrm{P}(\nu_{\mu}\to \nu_{\tau})=\sin^{2}2\theta \sin^{2}\frac{\Delta m^{2}_{12}L}{4E},~~ \mathrm{P}(\nu_{\mu}\to \nu_{\mu})=1-\sin^{2}2\theta \sin^{2}\frac{\Delta m^{2}_{12}L}{4E}.
\end{equation}
Atmospheric, solar and long baseline accelerator and reactor data are  described by the three-neutrino mixing. In this case
 transition probabilities depend on six parameters: atmospheric and solar mass-squared differences $\Delta m^{2}_{A}$ and  $\Delta m^{2}_{S}$, three mixing angles $\theta_{12}$, $\theta_{23}$, $\theta_{13}$ and one CP phase $\delta$.

Neutrino masses are usually labeled in such a way that $m_{1}$ and $m_{2}$ are connected  with solar neutrinos.
From the condition of the MSW resonance \cite{Wolfenstein:1977ue,Mikheev:1986wj}  follows that $\Delta m^{2}_{12}>0$. The solar mass squared difference is determined as follows $\Delta m^{2}_{S}=\Delta m^{2}_{12}$.

For $m_{3}$ there are two possibilities
\begin{enumerate}
  \item Normal ~ordering (NO)\quad   $ m_{3}>m_{2}>m_{1}$.
  \item Inverted~ ordering (IO)\quad   $m_{2}>m_{1}>m_{3}$.
\end{enumerate}
Atmospheric mass-squared difference can be determined as follows
\begin{equation}\label{Atm}
\Delta m^{2}_{A}=\Delta m^{2}_{23}~ (\mathrm{NO})\quad  \Delta m^{2}_{A}=|\Delta m^{2}_{13}|~(\mathrm{IO}).
\end{equation}
Using  (\ref{Atm}) from (\ref{Probabil2}) we can easily find that probabilities of $\nua{l}\to \nua{l'}$ transitions have the  form of the sum of atmospheric, solar and interference terms \cite{Bilenky:2015xwa}. In NO case we have
\begin{eqnarray}
&&P^{\mathrm{NO}}(\nua{l}\to \nua{l'})
=\delta_{l' l }
-4|U_{l 3}|^{2}(\delta_{l' l} - |U_{l' 3}|^{2})\sin^{2}\Delta_{A}\nonumber\\&&-4|U_{l 1}|^{2}(\delta_{l' l} - |U_{l' 1}|^{2})\sin^{2}\Delta_{S}
-8~[\mathrm{Re}~(U_{l' 3}U^{*}_{l 3}U^{*}_{l'
1}U_{l 1})\cos(\Delta_{A}+\Delta_{S})\nonumber\\
&&\pm ~\mathrm{Im}~(U_{l' 3}U^{*}_{l 3}U^{*}_{l'
1}U_{l 1})\sin(\Delta_{A}+\Delta_{S})]\sin\Delta_{A}\sin\Delta_{S},
\label{Genexp5}
\end{eqnarray}
For IO case we find
\begin{eqnarray}
&&P^{\mathrm{IO}}(\nua{l}\to \nua{l'})
=\delta_{l' l }
-4|U_{l 3}|^{2}(\delta_{l' l } - |U_{l' 3}|^{2})\sin^{2}\Delta_{A}\nonumber\\&&-4|U_{l 2}|^{2}(\delta_{l' l} - |U_{l' 2}|^{2})\sin^{2}\Delta_{S}
-8~[\mathrm{Re}~(U_{l' 3}U^{*}_{l 3}U^{*}_{l'
2}U_{l 2})\cos(\Delta_{A}+\Delta_{S})\nonumber\\
&&\mp ~\mathrm{Im}~(U_{l' 3}U^{*}_{l 3}U^{*}_{l'
2}U_{l 2})\sin(\Delta_{A}+\Delta_{S})]\sin\Delta_{A}\sin\Delta_{S},
\label{Genexp6}
\end{eqnarray}
where $\Delta_{A,S}=\frac{\Delta m^{2}_{A,S}L}{4E}$.
It is seen from comparison of these expressions that
$P^{\mathrm{NO}}(\nua{l}\to \nua{l'})$ and $P^{\mathrm{IO}}(\nua{l}\to \nua{l'})$ differ by the change of $U_{l(l') 1}\leftrightarrows U_{l(l') 2} $ and by the
sign of last terms.

In conclusion we present in the Table  \ref{tab:I} the result of the global analysis of existing neutrino oscillation data \cite{Esteban:2020cvm}.

\begin{table}
\caption{Values of neutrino oscillation parameters obtained from the global fit of existing data \cite{Esteban:2020cvm}}
\label{tab:I}
\begin{center}
\begin{tabular}{|c|c|c|}
  \hline  Parameter &  Normal Ordering& Inverted Ordering\\
\hline   $\sin^{2}\theta_{12}$& $0.310^{+0.013}_{-0.012}$& $0.310^{+0.013}_{-0.012}$
\\
\hline    $\sin^{2}\theta_{23}$& $0.582^{+0.015}_{-0.019}$& $ 0.582^{+0.015}_{-0.018}$
\\
\hline   $\sin^{2}\theta_{13}$ & $ 0.02240^{+0.00065}_{-0.00066}$&  $0.02263^{+0.00065}_{-0.00066}$
\\
\hline   $\delta $~(in $^{\circ}$) & $(217^{+40}_{-28})$& $ (280^{+25}_{-28})$
\\
\hline $\Delta m^{2}_{S}$& $(7.39^{+0.21}_{-0.20})\cdot 10^{-5}~\mathrm{eV}^{2}$&$(7.39^{+0.21}_{-0.20})\cdot 10^{-5}~\mathrm{eV}^{2}$\\
\hline $\Delta m^{2}_{A}$& $(2.525^{+0.033}_{-0.031})\cdot 10^{-3}~\mathrm{eV}^{2}$&$(2.512^{+0.034}_{-0.031})\cdot 10^{-3}~\mathrm{eV}^{2}$\\
\hline
\end{tabular}
\end{center}

\end{table}
The study of neutrino oscillations enter now into high-precision era. High precision measurements (at \% level) are necessary in order to solve such fundamental problems of neutrino physics as
\begin{itemize}
  \item What is the neutrino mass ordering?
  \item Is $CP$ is violated in the lepton sector and what is the precise value of CP phase $\delta$?
 \end{itemize}
Neutrino oscillation experiments allow to determine two neutrino mass-squared differences $\Delta m^{2}_{S}$ and $\Delta m^{2}_{A}$. The lightest neutrino mass and, correspondingly, absolute values of neutrino masses are at present unknown.

In recent tritium  KATRIN experiment \cite{Aker:2019uuj} the following bound was found
\begin{equation}\label{Katrin}
m_{\beta}=(\sum_{i}|U_{ei}|^{2}m^{2}_{i})^{1/2}<1.1~ \mathrm{eV}.    \end{equation}
From different recent cosmological measurements it was obtained \cite{Aghanim:2018eyx}
\begin{equation}\label{Cosmo}
\sum_{i}m_{i}<0.12~\mathrm{eV}.
\end{equation}

\section{On the Origin of Small Neutrino Masses}
Before starting the discussion of a possible origin of neutrino masses we would like to remind that particles with spin 1/2 can be Dirac or Majorana.

Dirac field $\psi(x)$ is a complex  (non hermitian) four-component field  which satisfies the Dirac equation. If a  Lagrangian is invariant under a global transformation $\psi(x)\to e^{i\Lambda}\psi(x)$ ($\Lambda$ is a constant) a charge is conserved and $\psi(x)$ is a field of  particles and antiparticles, which have opposite charges, same masses (due to the $CPT$ invariance)  and helicities $\pm 1$.

Majorana field  $\chi(x)$ is a two-component field which satisfies the Dirac equation and the  Majorana condition
\begin{equation}\label{Maj}
\chi(x)=\chi^{c}(x)=C\bar \chi^{T}(x),~~C\gamma^{T}_{\alpha}C^{-1}=-\gamma_{\alpha},~C^{T}=-C.   \end{equation}
There is no global invariance of a Lagrangian in the Majorana case. Majorana field is two-component field of truly neutral particles with helicities $\pm 1$.

Neutrino masses and mixing are generated by {\em a neutrino mass term}. The first neutrino mass term was proposed by V. Gribov and B. Pontecorvo \cite{Gribov:1968kq} in 1969. At that time it was established that the Lagrangian of the weak interaction had $V-A$ current $\times$ current form
\begin{equation}\label{Current}
    \mathcal{L}=-\frac{G_{F}}{\sqrt{2}}j^{CC}(j^{CC})^{\dag}
\end{equation}
where the leptonic current was given by the expression
\begin{equation}\label{Current1}
j^{CC,\mathrm{lep}}_{\alpha}
=2(\bar\nu_{eL}\gamma_{\alpha}e_{L}+\bar\nu_{\mu L}\gamma_{\alpha}\mu_{L}).
\end{equation}
Gribov and Pontecorvo  put themselves the following question: is it possible to introduce neutrino masses and mixing in the case if neutrino fields are left-handed $\nu_{eL}$, $\nu_{\mu L}$?\footnote{ It was a common opinion at that time that left-handed neutrinos are massless.}
 They understood that {\em if  the total lepton number $L=L_{e}+L_{\mu}$ is not conserved} it is possible to built a neutrino mass term in the case of $\nu_{eL}$, $\nu_{\mu L}$ fields. In fact, taking into account that the $C$-conjugated field
$ \nu_{lL}^{c}=C\bar\nu_{lL}^{T}$ is  right-handed, in the general three-flavor case (see \cite{Bilenky:1987ty}) we have the following unique mass term
\begin{equation}\label{Mjmass}
\mathcal{L}^{\mathrm{M}}(x)=-\frac{1}{2}\sum_{l',l}
\bar\nu_{l'L}(x)M^{\mathrm{M}}_{l'l}\nu_{lL}^{c}(x) +\mathrm{h.c.},\quad M^{\mathrm{M}}=(M^{\mathrm{M}})^{T}.
\end{equation}
The matrix  $M^{\mathrm{M}}$ can be diagonalized as follows
\begin{equation}\label{Mjmass1}
M^{\mathrm{M}}=U~m~U^{T}, \quad U^{\dag}~U=1,
\end{equation}
where $m_{ik}=m_{i}\delta_{ik},~ m_{i}>0$. From (\ref{Mjmass}) and (\ref{Mjmass1}) we find
\begin{equation}\label{Mjmass2}
\mathcal{L}^{\mathrm{M}}(x)=-\frac{1}{2}\sum^{3}_{i=1}m_{i}~
\bar\nu_{i}(x)\nu_{i}(x),
\end{equation}
where $\nu_{i}(x)$, the field of neutrino with mass $m_{i}$,
 satisfies the Majorana condition
\begin{equation}\label{Mj}
\nu_{i}(x)=\nu^{c}_{i}(x) =C\bar\nu^{T}_{i}(x).
\end{equation}
The flavor field $\nu_{lL}(x)$ is a "mixed" field
\begin{equation}\label{Mj1}
    \nu_{lL}(x) =\sum^{3}_{i=1}U_{li}\nu_{iL}(x),\quad (l=e,\mu,\tau)
\end{equation}
The mass term $\mathcal{L}^{\mathrm{M}}$ is called the Majorana mass term. It is the only possible mass term in which  the left-handed flavor fields  $\nu_{lL}$ enter. We would like to stress that in the framework of  purely phenomenological approach, we discussed, neutrino masses $m_{i}$ (and mixing matrix $U$)  are parameters. There are no any clues,  why neutrino masses are much smaller then lepton and quarks masses.

 Origin of neutrino masses and neutrino mixing is an open problem. Exist many different models. It is commonly  suggested that  {\em the Standard Model neutrinos are massless particles}.

 Masses of quarks and leptons are of the Standard Model origin.
 They are generated by  $SU_{L}(2)\times U_{Y}(1)$ invariant Yukawa interactions. In the case of leptons the Yukawa Lagrangian has the form
 \begin{equation}\label{Yukawa}
    \mathcal{L}^{Y}_{I}=-\sqrt{2}\sum_{l',l}\bar\psi_{l'L}
    Y_{l'l}l_{R}\phi+\mathrm{h.c.}
\end{equation}
Here
 \begin{equation}\label{doublets}
    \psi_{lL}=\left(
\begin{array}{c}
\nu_{lL} \\
l_L \\
\end{array}
\right)~~~(l=e,\mu,\tau),~~\phi=\left(
\begin{array}{c}
\phi_{+} \\
\phi_{0}\\
\end{array}
\right)
\end{equation}
  are lepton and Higgs doublets, $l_{R}$ is a  singlet, $Y$ is a dimensionless complex  matrix. After spontaneous symmetry breaking and diagonalization of $Y$ we come to the Dirac mass term
\begin{equation}\label{Dir}
\mathcal{L}^{Y}_{I}(x)=-\sum_{l=e,\mu,\tau}m_{l}~\bar l(x)~l(x).
\end{equation}
Here $l(x)=l_{L}(x)+l_{R}(x)$ is the Dirac field of leptons
$l^{-}$ ($Q=-1$ and antileptons  $ l^{+}$ ($Q=1$)
The lepton mass $m_{l}$ is given by the relation
\begin{equation}\label{lepmass}
m_{l}=y_{l}~v \quad (l=e,\mu,\tau).
\end{equation}
Here $y_{l}$ is a Yukawa coupling (eigenvalue of the matrix $Y$) and $v=(\sqrt{2}G_{F})^{-1/2}\simeq 246~\mathrm{GeV}$ is  the Higgs vev (electroweak scale).

All SM masses (masses of quarks, leptons, $W^{\pm}$ and $Z^{0}$ bosons, the mass of the Higgs boson) are proportional to $v$.\footnote{ This is connected with the fact that $v$ is the only parameter of the Standard Model which has a dimension $\mathrm{M}$.}

If neutrino masses are also of the Standard Model origin in this case
\begin{itemize}
  \item Right-handed singlets $\nu_{lR}$ enter into Lagrangian.
  \item The total lepton number is conserved and neutrinos with definite masses $\nu_{i}$ are Dirac particles.
\end{itemize}
Neutrino masses are given by the expression
\begin{equation}\label{Numass}
 m_{i}=y^{\nu}_{i}~v.
\end{equation}
Yukawa constants are determined by masses. For quarks and lepton  of the third  family we have
\begin{equation}\label{3fam}
y_{t}\simeq 7\cdot 10^{-1},~~ y_{b}\simeq 2\cdot 10^{-2},~~y_{\tau}\simeq 7\cdot 10^{-3}.
\end{equation}
Absolute values of neutrino masses are not known at present. However, assuming normal ordering of neutrino masses and using a conservative cosmological bound ($\sum_{i}m_{i}< 1$~eV) for the largest neutrino mass $m_{3}$ and neutrino Yukawa coupling $y^{\nu}_{3}$ we find the following bounds
\begin{equation}\label{bounds}
(5\cdot 10^{-2}\lesssim m_{3}\lesssim 3\cdot 10^{-1})~ \mathrm{eV},\quad
2\cdot 10^{-13}\lesssim    y^{\nu}_{3}\lesssim \cdot 10^{-12}.
\end{equation}
Yukawa couplings of quarks and lepton of the same family differ by about two orders of magnitude.  Neutrino Yukawa coupling $y^{\nu}_{3}$ is about ten orders of magnitude smaller then Yukawa couplings of top and bottom quarks and $\tau$-lepton.  It is very unlikely that {\em neutrino masses are of the same Standard Model origin as masses of leptons and quarks}. The Standard Model with left-handed, massless $\nu_{e}$, $\nu_{\mu}$, $\nu_{\tau}$ (without right-handed neutrino fields) is a minimal theory, originally proposed by Weinberg and Salam. We come to the conclusion that in order to generate small neutrino masses, observed in neutrino oscillation experiments, {\em we need a new beyond the Standard Model mechanism.}

A general method which allow to describe effects of a beyond the Standard Model physics is a method of the effective Lagrangian.
Effective Lagrangian is a dimension five or more non renormalizable  Lagrangian built from the Standard Model fields
and invariant under   $SU(2)_{L}\times U(1)_{Y}$  transformations.

Effective Lagrangians are generated by  beyond  the Standard model interactions of SM particles with heavy particles
 with masses much larger than $v$. In the electroweak region such interactions induce processes with virtual heavy particles, which are described by effective Lagrangians (fields of heavy particles are "integrated out"). Typical example  is the four-fermion, dimension six,  Fermi effective Lagrangian of the weak interaction.

In order to built an effective Lagrangian which generate a neutrino mass term , let us consider dimension $M^{5/2}$ $SU_{L}(2)\times U_{Y}(1)$  invariant
\begin{equation}\label{inv}
(\tilde{\phi }^{\dag}~ \psi_{lL}),
\end{equation}
where $\tilde{\phi }=i\tau_{2}\phi^{*}$ is a conjugated Higgs doublet. After spontaneous symmetry breaking  we have
\begin{equation}\label{inv1}
(\tilde{\phi }^{\dag}~ \psi_{lL})\to \frac{v}{\sqrt{2}}~\nu_{lL}.
\end{equation}
From (\ref{inv1}) it is obvious that (like in the Gribov-Pontecorvo case) we can built an effective Lagrangian which generates a neutrino mass term only if {\em the total lepton number is not conserved.} We come to the following unique expression for the effective Lagrangian (Weinberg \cite{Weinberg:1979sa})
\begin{equation}\label{Weinb}
\mathcal{L}_{I}^{\mathrm{W}}=-\frac{1}{\Lambda}~\sum_{l',l}
\overline{(\tilde{\phi }^{\dag} \psi_{l'L})}X_{l'l}(\tilde{\phi }^{\dag}~ \psi_{lL})^{c}+\mathrm{h.c.}
\end{equation}
Here $X$ is  $3\times 3$ dimensionless, symmetrical matrix and $\Lambda$ is a parameter which has dimension $M$ (the operator in $\mathcal{L}_{I}^{\mathrm{eff}}$ has a dimension $M^{5}$). The parameter $\Lambda$ characterizes a scale of a beyond the SM physics.

In connection with non conservation of $L$ by the Lagrangian (\ref{Weinb}) we would like to make the following general remark.
Global invariance and conservation of $L$ (and $B$) is not a fundamental symmetry of  QFT \cite{Weinberg:1980bf,Witten:2017hdv}.  Local gauge symmetry   ensure conservation of $L$ (and $B$) by the Standard Model Lagrangian. It is natural to expect that a beyond the Standard Model theory does not conserve $L$ (and $B$).

After spontaneous symmetry breaking from (\ref{Weinb}) we come to the Majorana mass term
\begin{equation}\label{Weinb1}
\mathcal{L}^{\mathrm{M}}= -\frac{1}{2}\,\sum_{l',l}
\bar\nu_{l'L}\,\frac{v^{2}}{\Lambda}~ X_{l'l}  ~\nu^{c}_{lL}+\mathrm{h.c.}=-\frac{1}{2}\sum^{3}_{i=1}m_{i}~\bar \nu_{i}\nu_{i}.
\end{equation}
Here $\nu_{i}= \nu^{c}_{i}$ is the field of the Majorana neutrino with the ``seesaw mass"
\begin{equation}\label{seemass}
m_{i}=\frac{v^{2}}{\Lambda}~x_{i}=\frac{v}{\Lambda}\cdot
(x_{i}v),
\end{equation}
where $x_{i}$ is the eigenvalue of the matrix $X$. In (\ref{seemass})  $(x_{i}v)$ is a ``typical" SM mass. Thus, the generation of neutrino masses via the effective Lagrangian mechanism leads to a suppression factor
\begin{equation}\label{seemass1}
\frac{v}{\Lambda}=\frac{\mathrm{EW~scale}}{\mathrm{scale~of~a~new~ physics}}.
\end{equation}
There are two unknown parameters ($x_{i}$ and $\Lambda$) in (\ref{seemass}). Thus, values of neutrino masses can not be predicted. However, if $\Lambda\gg v $ in this case Majorana neutrino masses $m_{i}$ are naturally much smaller than masses of leptons and quarks.

Notice that assuming $x_{3}\simeq 1$ (like Yukawa coupling of the top quark) for the scale of a new physics, responsible for neutrino masses, we find
\begin{equation}\label{scale}
\Lambda \simeq ( 10^{14}-10^{15})~\mathrm{GeV}.
\end{equation}

\section{On the Origin of the Weinberg Effective Lagrangian}
In this section we will briefly discuss a possible origin of the Weinberg effective Lagrangian (\ref{Weinb}). We will start with the simplest and most economical  scenario. Let us assume that
lepton-Higgs pairs interact with heavy Majorana leptons $N_{i}=N^{c}_{i}$~($i=1,2,..n$), $SU_{L}(2)$ singlets, via $SU_{L}(2)\times U_{Y}(1)$ interaction
\begin{equation}\label{heavy}
\mathcal{L}_{I}=-\sqrt{2}\sum_{l, i}(\bar \psi_{l L}\tilde{\phi })y_{li}~N_{iR}+\mathrm{h.c.}
\end{equation}
Here  $y_{li}$ are dimensionless constants.

{\em In the tree approximation} for low-energy processes with virtual heavy leptons at $Q^{2}\ll M^{2}_{i}$  we obtain the Weinberg effective Lagrangian in which
\begin{equation}\label{heavy1}
 \frac{1}{\Lambda}X_{l'l}=\sum^{n}_{i=1}y_{l'i}~\frac{1}{M_{i}}~y_{li}.
\end{equation}
Thus, masses of heavy Majorana leptons determine the scale $\Lambda$.

The mechanism, we discussed, is called the  type-I seesaw mechanism \cite{Minkowski:1977sc,GellMann:1980vs,Yanagida:1979as,
Glashow:1979nm,Mohapatra:1980yp}. Notice, that  the Weinberg effective Lagrangian can be also generated by interaction of heavy
triplet scalar bosons with a Higgs pair and lepton pair (type-II seesaw mechanism) and   by interaction of
lepton-Higgs pairs with heavy Majorana triplet leptons (type-III seesaw mechanism).

There exist numerous  {\em radiative neutrino mass models} which lead  to the Weinberg effective Lagrangian and, correspondingly, to the Majorana neutrino mass term.  In these models values of neutrino masses $m_{i}$ are suppressed by  loop mechanisms which require existence of different  beyond the Standard Model particles with masses which could be much smaller than $ 10^{15}$ GeV (see review \cite{Cai:2017jrq}).

\section{Conclusion}
In the first part of this report we considered a convenient phenomenology of neutrino oscillations in vacuum. In the second part  we discussed a possible (and plausible) origin of neutrino masses.

The approach, we considered, is based on the following
general assumptions
\begin{enumerate}
  \item There exist a beyond  Standard Model  interaction(s) of SM lepton and Higgs doublets and  new particles whose masses are  much larger than the electroweak scale $v$.
  \item Standard Model neutrinos  are massless left-handed particles.
 \end{enumerate}
 Beyond the SM interactions (after fields of heavy particles are integrated out) generate in the electroweak region an effective Lagrangian. From 2. it follows that independently on a type of model,  tree-level or radiative,  the only possible effective Lagrangian is $L$-violating, dimension five Weinberg Lagrangian which lead  to the most economical Majorana mass term.

The effective Lagrangian method of the generation of neutrino masses can explain (and, apparently, was inspired by) the smallness of neutrino masses. Values of neutrino masses $m_{i}$, neutrino mixing angles and $CP$ phase unknown parameters which depend on model and {\em can not be predicted.}

However, the following features are common for all models we discussed (in this sense are model independent)
\begin{enumerate}
  \item The number of neutrinos with definite masses $\nu_{i}$ is equal to the number of lepton flavors (three).
\item Neutrinos with definite masses $\nu_{i}$ are Majorana particles.
\end{enumerate}
Thus, the effective Lagrangian method of neutrino mass generation
predicts that there are {\em no transitions of flavor neutrinos into sterile states.} As it is well known, indications in favor of fourth (sterile) neutrino $m_{4}$ with mass in the range $(10^{-1}\lesssim m_{4}\lesssim 10)$ eV were obtained in different short baseline neutrino experiments. About 25 years ago in the accelerator LSND experiment indications in favor of  $\bar\nu_{\mu}\to \bar\nu_{e}$ were found. Later, these indications were confirmed by the MiniBooNE
accelerator experiment  in which transitions $\nua{\mu}\to \nua{e}$ were studied. The sterile neutrino anomaly was found also by reanalysis of old reactor neutrino  experiment data and by analysis of the data of GALLEX and SAGE  Gallium calibration experiments (see recent review \cite{Boser:2019rta}).

Several new short baseline reactor, accelerator, atmospheric and source neutrino experiments are going on or in preparations at present. From existing data it is not possible to make definite conclusions on the existence of sterile neutrinos (see talks presented at the NEUTRINO2020 conference http://nu2020.fnal.gov).

Notice, however, that recent combined analysis of the data of the reactor Daya Bay and Bugey-3 experiments and accelerator MINOS+  experiment, allows to exclude at 90 \% CL LSND and MiniBooNE allowed regions for $\Delta m^{2}_{14}< 5~\mathrm{eV}^{2}$  \cite{Adamson:2020jvo}, in new reactor DANSS experiment  \cite{Shitov} the best-fit point in the allowed region of previous reactor experiments is excluded at 5$\sigma$... etc.

The study of neutrinoless double $\beta$-decay  $(A,Z)\to (A,Z+2) +e^{-} +e^{-}$  is the most sensitive way which could allow us  to reveal the Majorana nature of neutrinos with definite masses $\nu_{i}$. In recent experiments the following lower limits on  half-lives of the $0\nu\beta\beta$-decay of different nuclei were reached: $T_{1/2}(^{76}\mathrm{Ge}) > 9\cdot 10^{25}$~yr~ (GERDA)  \cite{Agostini:2019hzm},
$T_{1/2}(^{136}\mathrm{Xe}^{136}) > 10.7\cdot 10^{25}$~yr ~(KamLAND-Zen) \cite{KamLAND-Zen:2016pfg}, $T_{1/2}(^{130}\mathrm{Te}) > 3.2\cdot 10^{25}$~yr~ (CUORE) \cite{Adams:2019jhp}.

 About one-two orders of magnitude larger half-lives are expected, if neutrinos are Majorana particles. In future $0\nu\beta\beta$- experiments such sensitivities are planned to be reached (see \cite{Detwiler}).

Summarizing, we discussed a plausible  (apparently, the most plausible) scenarios of  the origin of neutrino masses, based on such fundamental hypotheses  as a total lepton number violation by beyond the SM interactions. Crucial tests of this scenarios can be realized in experiments on
 \begin{itemize}
   \item The search for light sterile neutrinos.
   \item The search for neutrinoless double $\beta$-decay.
\end{itemize}

\end{document}